\documentclass[aps,twocolumn,a4paper,showpacs, superscriptaddress]{revtex4}
\usepackage[utf8]{inputenc}
\usepackage{amsmath,amssymb}
\usepackage{graphicx}
\usepackage{float}
\usepackage{color}

\newcommand{\bes}{\begin{subequations}}
\newcommand{\ees}{\end{subequations}}
\def\ben{\begin{eqnarray}}
\def\een{\end{eqnarray}}
\def\be{\begin{equation}}
\def\ee{\end{equation}}

\begin{document}
\title{Quest for eternal oscillons}

\author{T. S. Mendon\c ca}
\email{tiagobrouwer@msn.com}
\affiliation{Departamento de F\'{\i}sica Te\'orica - Instituto de F\'{\i}sica
	A. D. Tavares, Universidade do Estado do Rio de Janeiro, 
	R. S\~ao Francisco Xavier, 524. Rio de Janeiro, RJ, 20550-013, Brazil}
\author{H. P. de Oliveira}
\email{henrique.oliveira@uerj.br}
\affiliation{Departamento de F\'{\i}sica Te\'orica - Instituto de F\'{\i}sica
	A. D. Tavares, Universidade do Estado do Rio de Janeiro, 
	R. S\~ao Francisco Xavier, 524. Rio de Janeiro, RJ, 20550-013, Brazil}
\affiliation{Department of Physics and Astronomy, Bowdoin College, Brunswick, Maine, 04011, USA}


\begin{abstract}
Nonlinear field theories produce unstable but long-lived configurations known as oscillons. These structures have been studied with asymmetric and symmetric double-well potentials and extended to other forms of potentials. In the present work, we examine the consequences of considering higher-order field theories, where we have used a generalization of the symmetric double-well potential and a $\phi^6$ potential. Consequently, we have found  $3+1$ spherically symmetric oscillons with significantly large lifetimes without parameter fine-tuning.
\end{abstract}
\maketitle

%%%%%%%%%%%%%%%%%%

\section{Introduction}

The nonlinear aspect present in the equations governing some field theories poses immense challenges for investigating their physical consequences. As a relevant illustrative example, we mention a nonlinear wave equation arising from the Klein-Gordon equation with a single real scalar field with a symmetrical double-well potential of the $\phi^4$ model.  The simple case of $1+1$ dimensions reveals a nontrivial nonstatic solution known as boosted topological defects or boosted kinks \cite{rajaraman,vachaspati}. The dynamics of the collision of a kink and an antikink discloses complex dynamics from which the initial velocity kinks' velocity plays a crucial role for outcomes \cite{campbell_83,aninos_oliveira_91,shnir_18}. 

We can go a step further in considering the case of $3+1$ spherically symmetric Klein-Gordon equation in the $\phi^4$ model. In this case, the Derrick theorem prevents the existence of static solutions \cite{derrick}. Nevertheless, further studies of the scalar field dynamics uncovered unstable oscillating but long-lived structures known as oscillons \cite{pulson,gleiser_94}. These structures can be relevant in distinct areas of physics varying from optics to cosmology \cite{lozanov,farhi}, with the caveat the duration of oscillons be greater than the dynamic scales of these systems.

One of the main aspects that make oscillons attractive is their long duration. It is a phase achieved after the scalar field collapse, releases a large amount of its energy, and then starts to oscillate almost without radiating energy away. By using an innovative numerical treatment, Honda and Choptuik \cite{honda_choptuik} have shown evidence that oscillons can live forever. Like most current oscillation simulations, they considered an initial Gaussian scalar field profile with a free radial parameter $r_0$. Depending on the value of $r_0$, the oscillon phase can live longer. More interestingly, there are resonant peaks in the oscillon lifetime when viewed in the function of $r_0$. It means that by fine-tuning $r_0$, it is possible to generate oscillons with arbitrarily long lifetimes. The resonant mountain, as referred to by Gleiser and Krackow \cite{gleiser_krackow_20}, was later confirmed by them \cite{gk_31}.   

In the present manuscript, we investigate the existence of long-lived oscillons considering potentials that can be viewed as direct generalizations of the double-well potential \cite{gleiser_94,gleiser_02,honda_choptuik,fodor_06}. Nevertheless, we mention the existence of highly long-lived oscillons associated with other potentials such as the sine-Gordon and some classes of convex potentials \cite{zhang, salmi} and monodromy potentials (see Refs. \cite{olle} and references therein).  

%Most of the works on oscillons consider symmetric or asymmetric double-well potentials \cite{gleiser_94,gleiser_02,honda_choptuik,fodor_06}. In the present manuscript, we investigate the existence of long-lived oscillons considering different potentials. The first is a generalization of the $\phi^4$ symmetric double-well potential given by
%
The first potential is a generalization of the $\phi^4$ symmetric double-well potential given by
\begin{eqnarray}
V(\phi) = \frac{1}{4}(\phi_{vac}^{2n}-\phi^{2n})^2, \label{eq1} 
\end{eqnarray}

\noindent where $\phi_{vac}$ is the asymptotic vacuum value of the scalar field, $n \geq 1$ is a integer number with $n = 1$ recovering the potential of the $\phi^4$ model. This potential describes a generalization of the $(1+1)$ dimensional kinks known as the compactons \cite{bazeia_14}. The collision of compactons was studied by Bazeia et al. \cite{mendonca_compactons} exhibiting a wide variety of outcomes. The formation of long-lived oscillatory structures at rest or moving appears not to be limited to small impact velocities. The second potential we have chosen belong to the Christ-Lee model \cite{christ_lee} or the parametric $\phi^6$ model \cite{saxena}: %\cite{christ_lee} with the following form considered by Saxena et %al.\cite{saxena}:
\begin{eqnarray}
V(\phi) = \frac{1}{8(1+\epsilon^2)}(\epsilon^2+\phi^2)(\phi_{vac}^2-\phi^2)^2, \label{eq2}
\end{eqnarray}

\noindent where $\epsilon$ is a free parameter.  

We organize the paper as follows. In Section II, we present the Klein-Gordon equation briefly and describe the numerical method we have adopted. It is based on the Galerkin-Collocation method \cite{hpo_cqg_recent}  that consists in adopting features of both Galerkin and Collocation methods. In particular, we have defined a set of basis functions that satisfy exactly the boundary conditions where we consider the whole radial domain. We present the code validation in Section III with the convergence of the scalar field total energy and the reproduction of the oscillon phase already established \cite{gleiser_95}. In Section IV, we perform numerical simulations with the potentials (\ref{eq1}) and (\ref{eq2}). We have found convincing numerical evidence of extremely long-lived oscillons in both cases. Finally, in Section V we conclude.

%########################################################
\section{The numerical method: brief descritption}
%########################################################

The action for a self-interacting scalar field in $3+1$ dimensions is 
\begin{eqnarray}
S[\phi]=\int\,d^4x\,\left(\frac{1}{2} \partial_\mu \phi \partial^\mu \phi - V(\phi)\right) \label{eq3}
\end{eqnarray}

\noindent where $V(\phi)$ is the potential. We can derive the nonlinear Klein-Gordon equation straightforwardly from the above action and present it in the following useful form:
\begin{eqnarray}
\phi_{,t} &=& \Pi \label{eq4}\\
\nonumber \\
\Pi_{,t} &=&  \frac{2}{r}\phi_{,r} + \phi_{,rr} - \frac{\partial V(\phi)}{\partial \phi}, \label{eq5}
\end{eqnarray}

\noindent where $\Pi=\Pi(t,r)$ and $\phi=\phi(t,r)$. Here, we shall consider the dynamics generated by the compact and the Chris-Lee potentials given by Eqs. (\ref{eq1}) and (\ref{eq2}), respectively, but the numerical method described in the sequence is valid for any potential function.
% As we have mentioned previously, for the current studies of oscillons, the potential $V(\phi)$ is in general chosen to be symmetric or asymmetric double-well potentials. Here, we shall consider the dynamics generated by the compact and the Chris-Lee potentials given by Eqs. (\ref{eq1}) and (\ref{eq2}), respectively, but the numerical method described in the sequence is valid for any potential function.

We briefly describe the numerical spectral method employed to integrate the wave equation expressed by Eqs. (\ref{eq4}) and (\ref{eq5}). There are two issues in this wave equation that are decisive for implementing the numerical algorithm. The first is the $1/r$ term that must be regular when calculated at the origin, and the second is that the scalar field is an even parity with respect to $r$. Then, the expansion of the scalar field near the origin reads as
\begin{eqnarray}
\phi(t,r) = \phi_0(t)+\phi_2(t)r^2 + \mathcal{O}(r^4). \label{eq6}
\end{eqnarray}

\noindent Also, for $r \rightarrow \infty$, the scalar field assumes its vacuum value, i. e. $\phi_\infty =\phi_{\mathrm{vac}}$. To guarantee the finiteness of the scalar field energy, it is necessary that asymptotically that
\begin{eqnarray}
\phi(t,r) -\phi_{\mathrm{vac}} \leq \mathcal{O}(r^{-2}). \label{eq7}
\end{eqnarray}

\noindent  Similar conditions are valid for the field $\Pi(t,r)$.
	
The next step is to establish spectral approximations for the scalar field $\phi$ and the function $\Pi$ to satisfy the conditions above. Then, we can write 
\begin{eqnarray}
\phi_N(t,r) &=& \phi_{\mathrm{vac}}+\sum_{k=0}^N\,\hat{\phi}_{2 k}(t) \psi_{2 k}(r), \label{eq8}\\%\frac{1}{2}(SB_{2 k+2}(r)-SB_{2 k}(r)), \\
\nonumber \\
\Pi_N(t,r) &=& \sum_{k=0}^N\,\hat{\Pi}_{2 k}(t) \psi_{2 k}(r), \label{eq9}
\end{eqnarray}

\noindent where $N$ is the truncation order that dictates the number of unknown modes $\hat{\phi}_{2 k}(t)$ and $\hat{\Pi}_{2 k}(t)$ related to the fields $\phi$ and $\Pi$, respectively. We demand that the basis functions $\psi_{2 k}(r),\,k=0,1,..N$ have even parity functions with respect to the origin and decay according to the relation (\ref{eq7}). We establish the basis functions with the polynomials $SB_k(r)$ defined in Boyd's book \cite{boyd}
\begin{eqnarray}
SB_k(r) = \sin\bigg[(k+1)\mathrm{arcot}\left(\frac{r}{L_0}\right)\bigg], \label{eq10}
\end{eqnarray}

\noindent where $L_0$ is the map parameter. The above polynomials were employed for spectral codes to generate Brill wave initial data \cite{deol_rod_12} and later in Refs. \cite{ledvinka, alcoforado_bssn} for dynamical problems in general relativity. The even polynomials have the even parity about the origin, and by requiring the appropriate asymptotic behavior, we define 
\begin{eqnarray}
\psi_{2 k}(r)= \frac{(2k+1)}{2k+3}SB_{2k+2}(r)-SB_{2k}(r), \label{eq11}
\end{eqnarray}

\noindent for all $k=0,1,..,N$, and one may show that  $\psi_{2 k}(r) \sim \mathcal{O}(r^ {-3})$.

We substitute the spectral approximations (\ref{eq8}) and (\ref{eq9}) into Eqs. (\ref{eq4}) and (\ref{eq5}) to form the corresponding residual equations given by
\begin{eqnarray}
\mathrm{Res}_\phi(t,r) &=& \phi_{N,t}-\Pi_{N} \label{eq12}\\ 
\nonumber \\
\mathrm{Res}_\Pi(t,r) &=& \Pi_{N,t} - \phi_{N,rr} - \frac{2}{r} \phi_{N,r} + \frac{\partial V(\phi)}{\partial \phi}.  \label{eq13}
\end{eqnarray}

\noindent The above residual equations do not vanish due to the assumed approximations. According to the Collocation method, we impose the residual equation to vanish at a specific set of $N+1$  collocation points designated by $r_j$, with $j=1,2.., N+1$. In this case, the residual equations become the following dynamical system
\begin{eqnarray}
\phi_{j,t} &=& \Pi_j, \label{eq14} \\
\nonumber \\
\Pi_{j,t} &=& (\phi_{,rr})_j + 2 \left(\frac{\phi_{,r}}{r}\right)_j - \left(\frac{\partial V(\phi)}{\partial \phi}\right)_j. \label{eq15}  
\end{eqnarray}

%\noindent where the subscript $j$ indicates that the quantities are evaluated at the collocation points $r_j$, for instance, 

\noindent The subscript $j$ indicates that the quantities are evaluated at the collocation points $r_j$. For instance, the values of the scalar field $\phi_j(t)$ 

\begin{eqnarray}
\phi_j(t)=\phi_N(t,r_j)=\phi_{vac} + \sum_{k=0}^N\,\hat{\phi}_{2k}(t)\psi_{2k}(r_j), \label{eq16}
\end{eqnarray}

\noindent for all $j=1,2,..,N+1$. As a consequence, the values $\phi_j(t)$ are related to the corresponding modes $\hat{\phi}_{2k}(t)$. Both sets of values and modes constitute the physical and the spectral representations of the scalar field. In particular, we use the spectral respresentation to calculate the $1/r$ term on the RHS of Eq. (15) at the origin. We obtain similar relations for the values present in Eqs. (14) and (15) as

\[\Pi_j=\Pi_N(t,r_j),\;(\phi_{,rr})_j =\left(\frac{\partial^2 \phi_N}{\partial r^2}\right)_{r_j},\]

\noindent and so forth. %It is important to note that there is a relation between the values $\phi_j(t),\;j=1,2,.., N+1$ and the modes $\hat{\phi}_k,\;k=0,1,.., N$,  with both quantities constituting the physical and the spectral representations of the scalar field. In particular, we use the spectral representation to calculate the $1/r$ term on the RHS of Eq. (\ref{eq15}) at the origin. The choice of the basis functions guarantees the regularity of this term at $r=0$.

We define now the set of $N+1$ collocation points $r_j$. First, we introduce a computational domain spanned by $-1 \leq x \leq 1$ such that
\begin{eqnarray}
r = \frac{L_0 x}{\sqrt{1-x^2}}. \label{eq17}
\end{eqnarray}

\noindent It means that we have extended the physical domain to include negative values of $r$ or $-\infty < r < \infty$. In the sequence, we choose the Chebyshev-Gauss-Lobatto points $x_j$ as
\begin{eqnarray}
x_j=\cos\left(\frac{\pi j}{2(N+1)}\right), \label{eq18}
\end{eqnarray}

\noindent where $j=0,1,..,2(N+1)$. From these points, we select $N+1$ collocation points in the region $0 \leq r < \infty$ by setting $j=1,2,.., N+1$. Notice that the point $r_{N+1}$ corresponds to the origin. 

The scheme of integration is described as follows. From the initial data $\phi(0,t)=\phi_0(r)$ and $\Pi(0,t)=\Pi_0(r)$, we determine the initial values $\phi_j(0),\,\Pi_j(0)$ as well the initial modes $\hat{\phi}_k(0)$. In the sequence, we calculate the initial values of $\phi_{j,t}$ and $\Pi_{j,t}$ from Eqs. (\ref{eq14}) and (\ref{eq15}), respectively, allowing the determination of the values $\phi_j,\,\Pi_j$ in the next time step. We repeat the process and providing the scalar field evolution. We remark that in the specific numerical investigation of oscillons, it is necessary to perform integration for long times, typically until $t \gtrsim 10^4$. It requires the introduction of some sort of dissipation to absorb the scalar field radiation at large distances. We have added the dissipation term term $-\gamma_0(r) \Pi$ on the RHS of Eq. (5) with the coefficient $\gamma_0$ given by
\begin{eqnarray}
%\gamma_0(r) =\begin{cases}
%0,& \mathrm{if}\;\;r < r_{abs} \\
%0.05,& \mathrm{if}\;\; r \geq r_{abs}, \label{eq18}
%\end{cases}
\gamma_0(r) = \frac{1}{2}K_0^2\left[1+\tanh(r-r_{abs})\right], \label{eq19}
\end{eqnarray}

\noindent where $r_{abs}$ indicates the starting point of the damping term. As we are going to see next, $r_{abs}$ has to be larger than the bubble's core.

%%%%%%%%%%%%%%%%%%%%%%%%%%%%%%%%%
\section{Validating the code}
%%%%%%%%%%%%%%%%%%%%%%%%%%%%%%%%%

In all numerical tests, we have considered the initial data $\Pi_0=\Pi(0,r)=0$, and $\phi_0(r) = \phi(0,r)$ as a Gaussian-shaped function used for most of the works on oscillons: 
\begin{equation}
\phi_0(r)=(\phi_c-\phi_{vac}) \mathrm{e}^{-r^2/r_0^2} + \phi_{vac}, \label{eq20}
\end{equation}

\noindent where  $\phi_c$ and $\phi_{vac}$ are scalar field values at the bubble's core, and the global vacuum; $r_0$ is a control parameter. By fixing $\phi_c$ and $\phi_{vac}$, the long-lived oscillons emerge in the interval $R_{\mathrm{min}} \leq r_0 \leq R_{\mathrm{max}}$ depending on the potential $V(\phi)$ under consideration. In what follows, we have fixed $\phi_c=1$ and $\phi_{vac}=-1$ and performed all numerical experiments with the Cash-Karp adaptive stesize integrator \cite{cash_karp}.

The first numerical test is to verify the balance of the scalar field energy due to the presence of the dissipative term. The total energy of the scalar field is 
\begin{eqnarray}
E = 4\pi\,\int_0^\infty\,\left(\frac{1}{2}\Pi^2+\frac{1}{2}\phi_{,r}^2 + V(\phi)\right)r^2\,dr. \label{eq21}
\end{eqnarray}

\noindent Here, we assume that $V(\phi)$ is the symmetric double-well potential given by

\begin{eqnarray}
V(\phi) = \frac{1}{4}(\phi^2-\phi_{vac}^2)^2. \label{eq22}
\end{eqnarray}

By considering the dissipative term (\ref{eq19}), the total energy $E$ is no longer conserved but satisfies the following exact balance equation

\begin{eqnarray}
C(t) \equiv \frac{dE}{dt} + 4\pi\,\int_0^\infty \gamma_0(r)\, \Pi^2 r^2 dr = 0,\label{eq23}
\end{eqnarray}

\noindent where we introduce the function $C(t)$ to indicate the error of the numerical solution. We proceed by calculating both terms arising from the spectral algorithm with increasing resolution and selecting the maximum deviation $|C_{\mathrm{max}}|$. In the Appendix, we present the details of the approximations for calculating the total energy $E$ and the integral involving the dissipative term.

For the sake of illustration, we present in Fig. 1(a) the plots of $\frac{dE}{dt}$ and the integral of the dissipative term present in Eq. (22). We have fixed the following parameters: $r_0=2.5,\,K_0=0.5,\,r_{abs}=30,$ map parameter $L_0=30$, and the truncation order $N=200$ with the integration until $t=2,000$. And Fig. 1(b) shows a clear exponential decay of the maximum deviation $|C_{\mathrm{max}}|$ with the increase of the resolution characterized by the truncation order $N$. Notice that we have obtained a maximum deviation of order of $\mathcal{O}(10^{-9})$ for $N=350$.

\begin{figure}[htb]
	\includegraphics[width=7.cm,height=6cm]{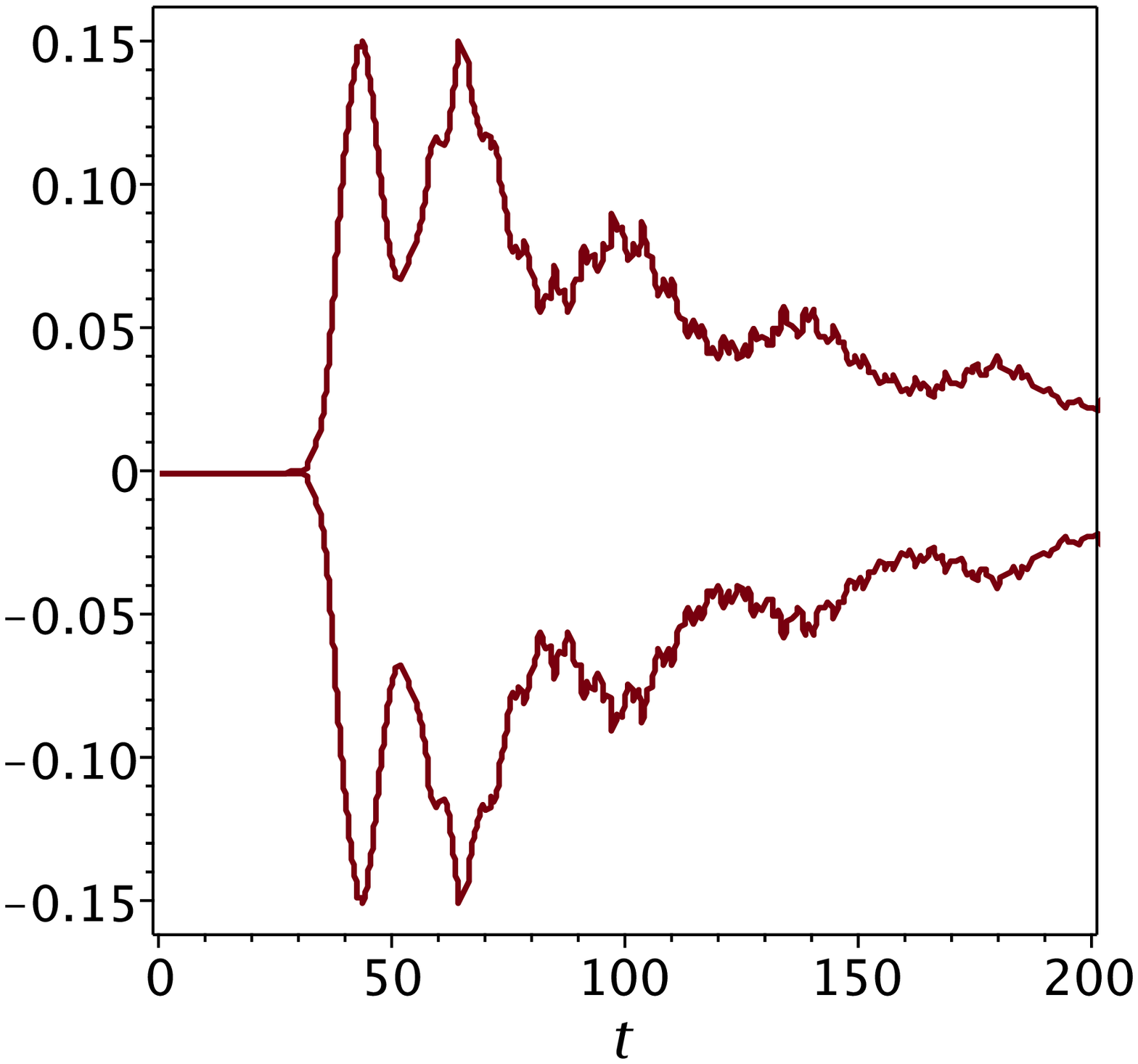}
	\includegraphics[width=7.cm,height=6cm]{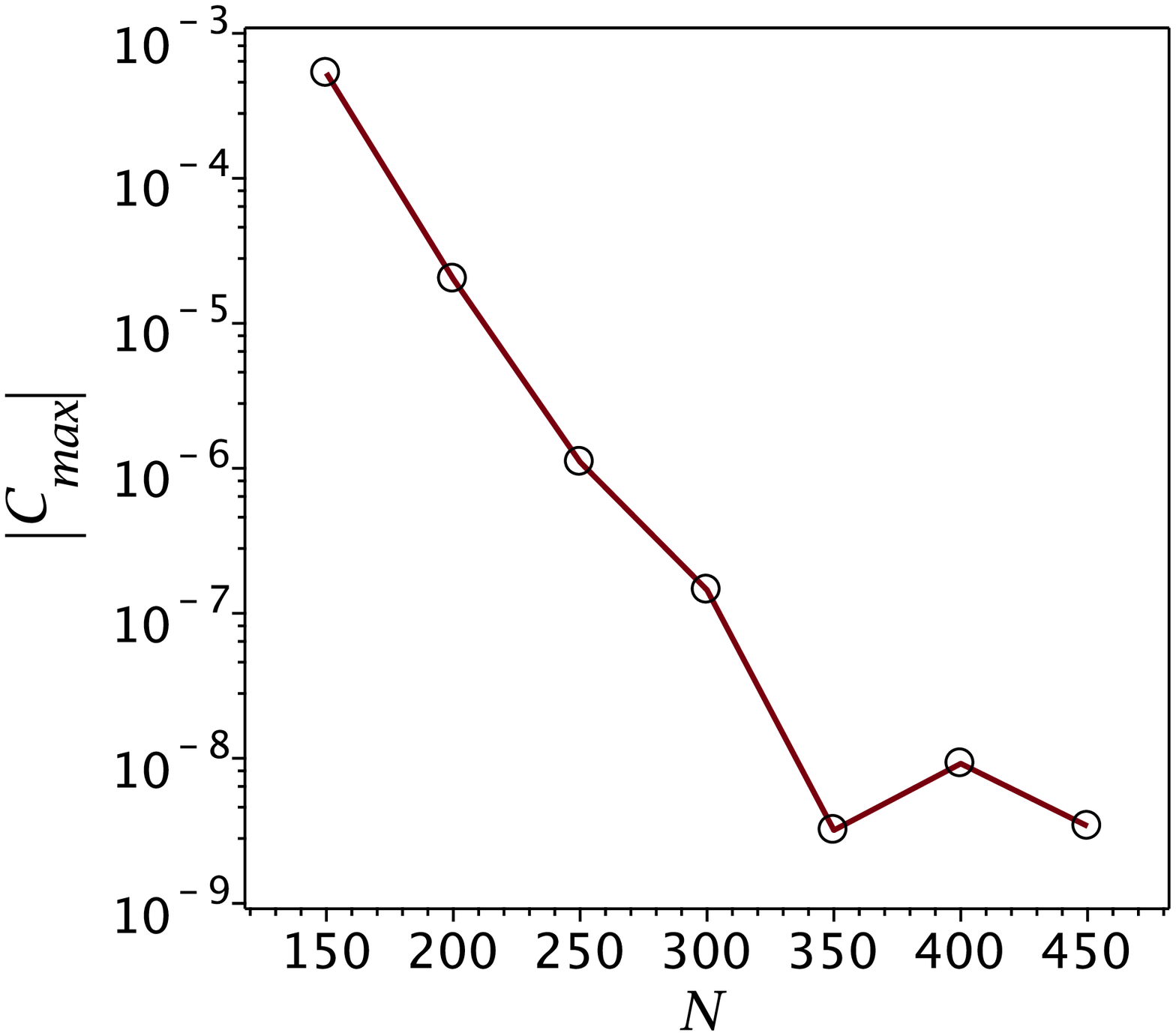}
	\caption{Upper panel: Samples of $dE/dt < 0$ and the integral of the dissipation term given by the second term on the RHS Eq. (22). Lower panel: log-plot of the maximal devitation, $|C_{max}|$, versus the truncation order $N$ showing a clear exponential decay. We have set $r_0=2.5$ and map parameter $L_0=30$. The final time of integration is $t_f=2,000$.} 
\end{figure}

\begin{figure}[htb]
	\includegraphics[width=6.5cm,height=5.5cm]{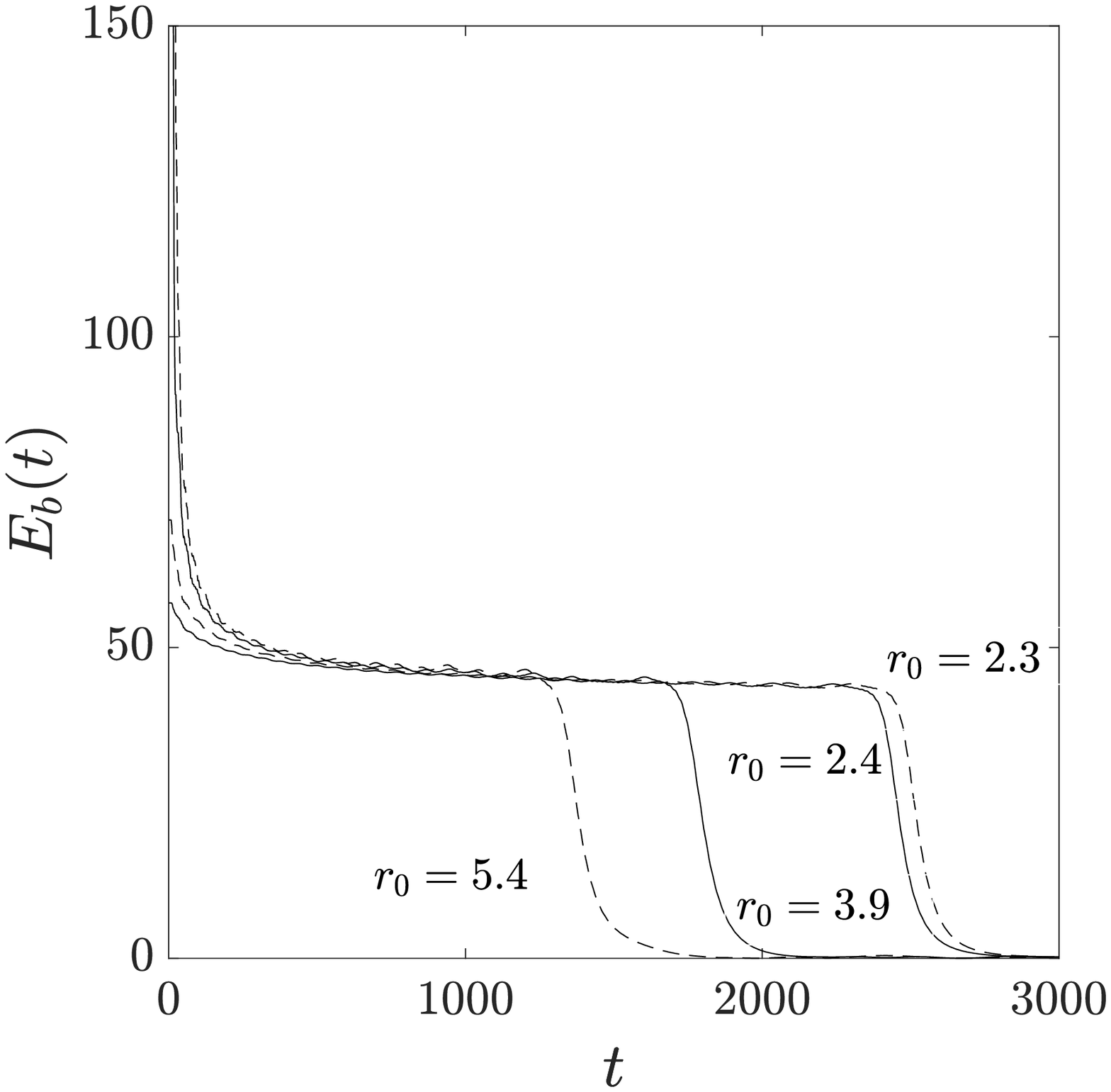}
	\includegraphics[width=6.5cm,height=5.5cm]{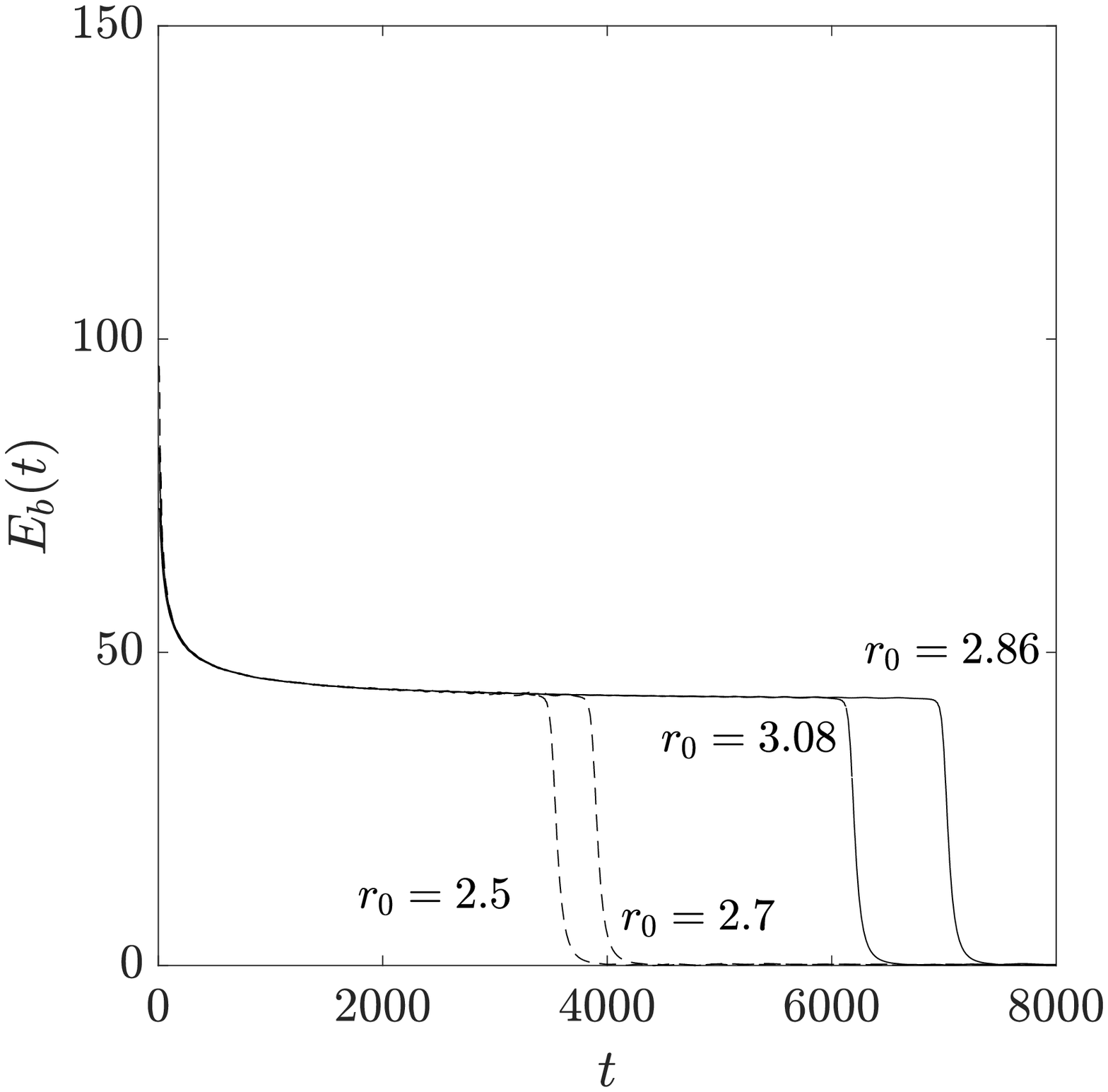}
	\caption{Plots of the bubbles' energy versus time for several values of $r_0$. The continuous and dashed lines correspond to the Gaussian and tanh initial configurations, respectively.} 
\end{figure}

\begin{figure}[htb]
	\includegraphics[width=6.5cm,height=5.5cm]{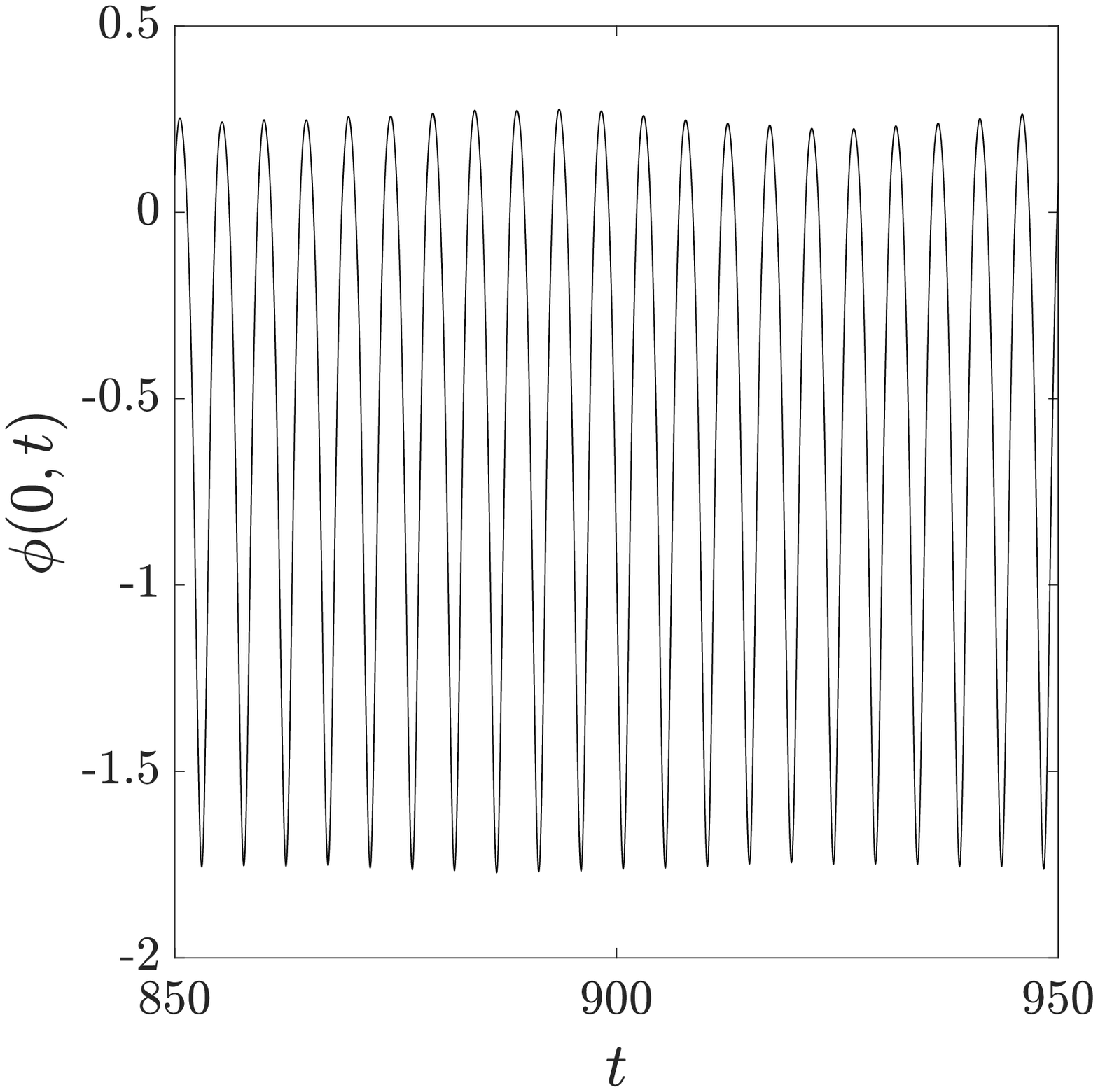}
	\includegraphics[width=6.5cm,height=5.5cm]{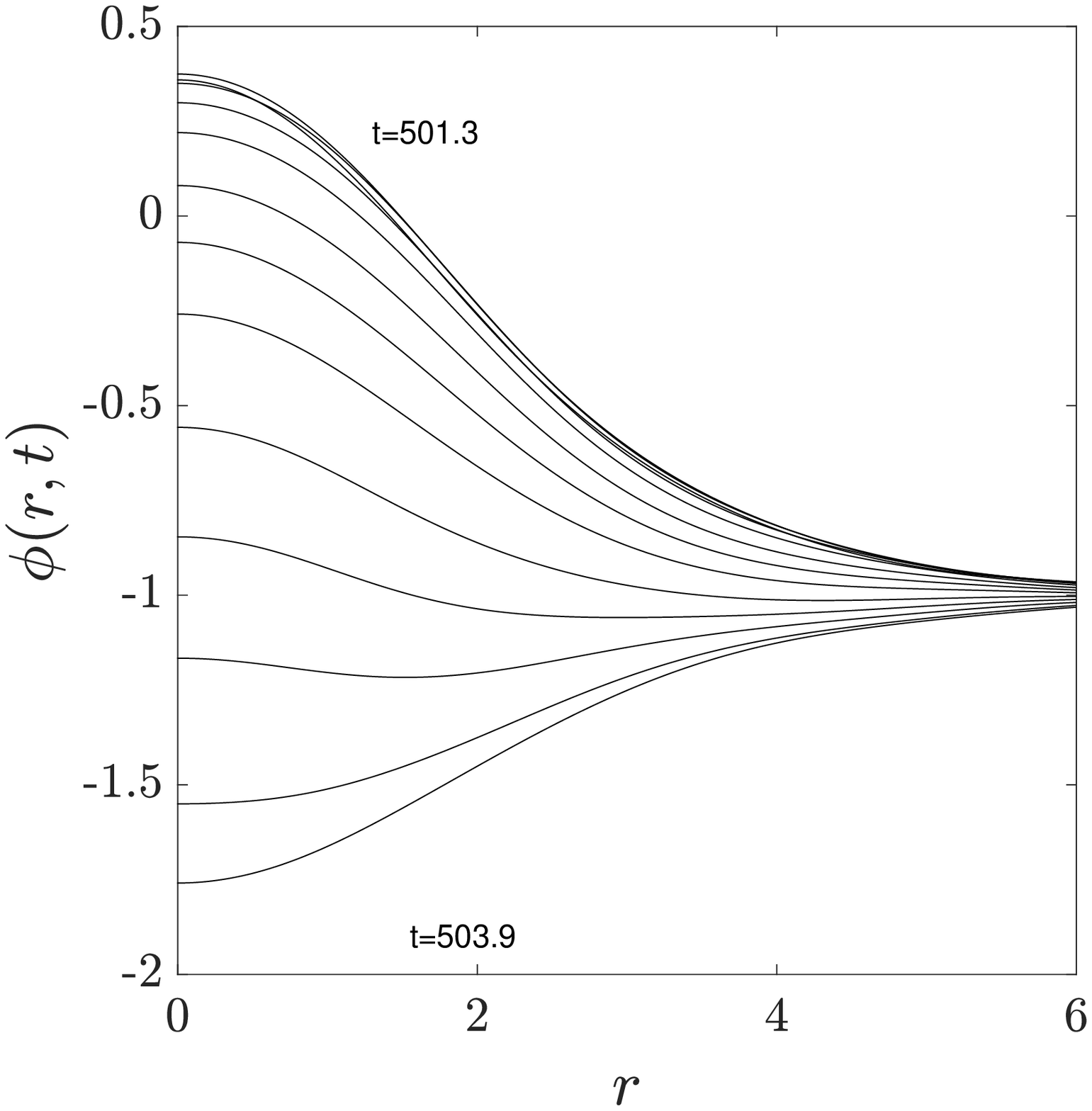}
	\caption{Upper panel: time evolution of the scalar field at the origin or the bubble's core during the oscillon stage for the Gaussian initial configuration with $r_0=2.7$. Lower panel: snaphosts of the oscillon with $r_0=2.5$. As in Ref. \cite{gleiser_95}, the snapshots are $\Delta t = 0.2$ apart.} 
\end{figure}

The remaining tests are qualitative in the sense of reproducing some of the results found in the Copeland, Gleiser, and Muller (CGM) \cite{gleiser_95} paper using the present numerical algorithm. In all numerical experiments, we have set a resolution of $N=250$, map parameter $L_0=15$, and introduced the dissipation $\gamma_0(r)$ given by Eq. (\ref{eq18}). 

As pointed out by CGM \cite{gleiser_95}, the Gaussian bubbles are sensitive to the value of $r_0$ with fixed $\phi_c=-\phi_{\mathrm{vac}}=1$ to achieve the stage of oscillon. Accordingly, in the interval, $2.4 \lesssim r_0 \lesssim 4.5$, the scalar field undergoes a stage of the oscillatory, long-lived stability period with almost no energy radiated away. This oscillatory stage is known as the oscillon stage and has a duration of about $10^3 - 10^4$. The range of $r_0$ depends on the initial profile of the bubble, but the subsequent oscillatory stage or the generated oscillon is the same.

\begin{figure}[htb]
	\includegraphics[width=7.5cm,height=6.5cm]{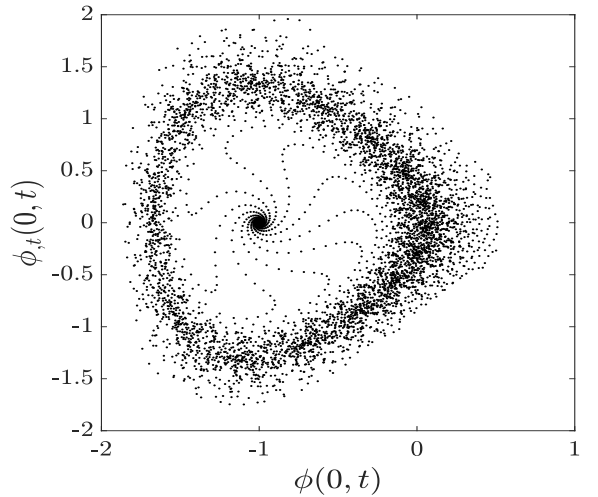}
	\caption{Phase space portrait of an oscillon stage for $r_0=3.0$. The sampling is taken at each $\Delta t =1$.} 
\end{figure}

We present in Fig. 2 the bubble's total energy defined by
\begin{eqnarray}
E_b(t) = 4\pi\,\int_0^{r_s}\,\left(\frac{1}{2}\Pi^2+\frac{1}{2}\phi_{,r}^2 + V(\phi)\right)r^2\,dr, \label{eq24}
\end{eqnarray}

\noindent for several values of $r_0$, and taking into account an additional initial data $\phi_0(r) = -\tanh(r-r_0)$ . Here, $r_s \gg r_0$, and by following CGM \cite{gleiser_95} we have set $r_s=10$ and $15$ for the Gaussian and tanh bubbles, respectively. It is remarkable to notice that after the bubbles released a fraction of their energy, the oscillon stage emerges with nearly constant energy regardless of the initial configuration and the initial radius $r_0$. In the panels of Fig. 2, the continuous and dashed lines correspond to the Gaussian and tanh bubbles, respectively. 

Fig. 3 shows another qualitative validation of our numerical approach. We exhibit in Fig. 3(a) the evolution of the scalar field at the origin, $\phi(0,t)$, during the oscillon stage with $r_0=2.7$. In Fig. 3(b), we have a sequence of snapshots of the oscillon with $r_0=2.5$. These figures are in agreement with Fig. 4 found in Ref. \cite{gleiser_95}.

An advantageous way of visualizing the evolution of the oscillon at the origin $r=0$ is using the phase portrait as shown in Fig. 4. We have considered the time interval $\Delta t= 1.0$ and $r_0=3.0$ according to Ref. \cite{gleiser_95}. At the end of evolution, the spiraling indicates the collapse of the oscillon into $\phi_{vac}=-1$. The total time of integration was $t_f=6000$. 

%%%%%%%%%%%%%%%%%%%%%%%%%%%%%%%%%%%%%%%%%
\section{(Very) Long lived oscillons}
%%%%%%%%%%%%%%%%%%%%%%%%%%%%%%%%%%%%%%%%%

%In this Section we have performed numerical simulations to achieve the oscillon phase starting from the initial Gaussian profile (\ref{eq19}) with $\phi_c=-\phi_{vac}=1$ leaving $r_0$ as a free parameter, but with the new potentials (\ref{eq1}) and (\ref{eq2}). We have set the resolution characterized by $N=400$, map parameter $L_0=20$, and the dissipation factor $\gamma_0(r)$ fixed as shown in Eq. (\ref{eq18}), where $r_{abs}=70$. We have tested different resolutions up to $N=1,000$, and locations of the dissipation provided that $r_{abs} \gg r_0$, without considerable change of the results.

In this Section we have performed numerical simulations to achieve the oscillon phase starting from the initial Gaussian profile (\ref{eq19}) with $\phi_c=-\phi_{vac}=1$ leaving $r_0$ as a free parameter, but with the new potentials (\ref{eq1}) and (\ref{eq2}). We have set the minimum and maximal resolutions characterized by $N=600$ and $1,000$, respectively, map parameter $L_0=20$, and the dissipation factor $\gamma_0(r)$ given by Eq. (\ref{eq19}), where $K_0=0.15$ and $r_{abs} \geq 10 r_0$. In all simulations, we have monitored the error in the energy balance measured by $C(t)$ (cf. Eq. (\ref{eq23})) such that the maximal deviation is of order $\mathcal{O}(10^{-4})$. We have noticed that by increasing $n$ in the potential (\ref{eq1}), we increase the order of the nonlinear terms, which requires a better resolution to achieve the maximal deviation in the energy balance.

In the sequence, we have focused on characterizing the long-lived oscillon states mainly through the plots of the bubble energy given by Eq. (\ref{eq24}), where $r_s=15$ in all numerical experiments.  %We have considered the potentials provided by Eqs. (3) and (4).

\begin{figure}[htb]
	\includegraphics[width=6.5cm,height=5.5cm]{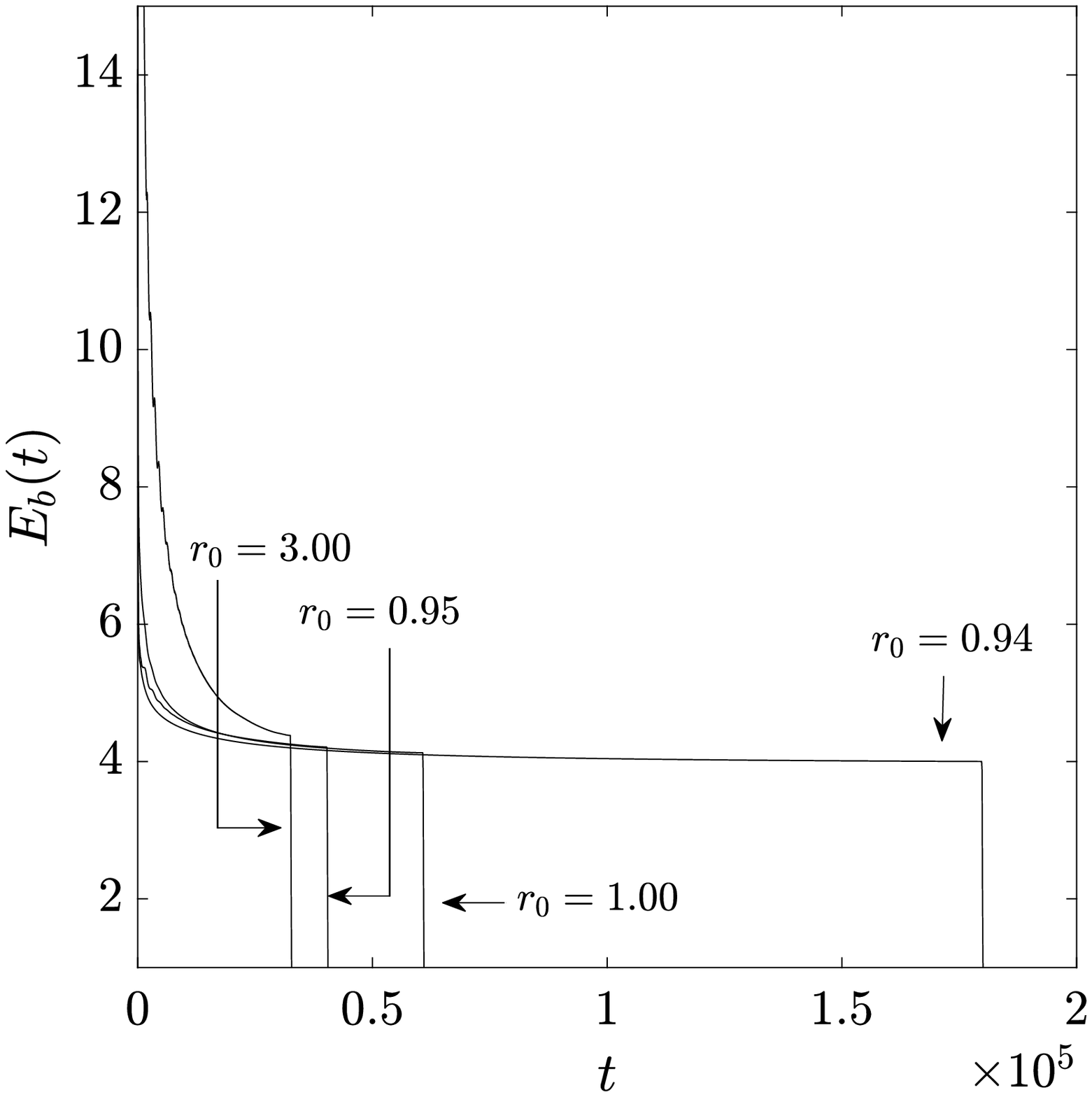}
	\caption{Time evolution of the bubble energy, $E_b(t)$ (cf. Eq. (\ref{eq24})), for the compacton potential (\ref{eq1}) for $n=2$ and distinct values of the bubble's radius $r_0$. We have used the resolution $N=600$ and $r_{abs}=30$.} 
\end{figure}

\begin{figure}[htb]
    \includegraphics[width=6.5cm,height=5.5cm]{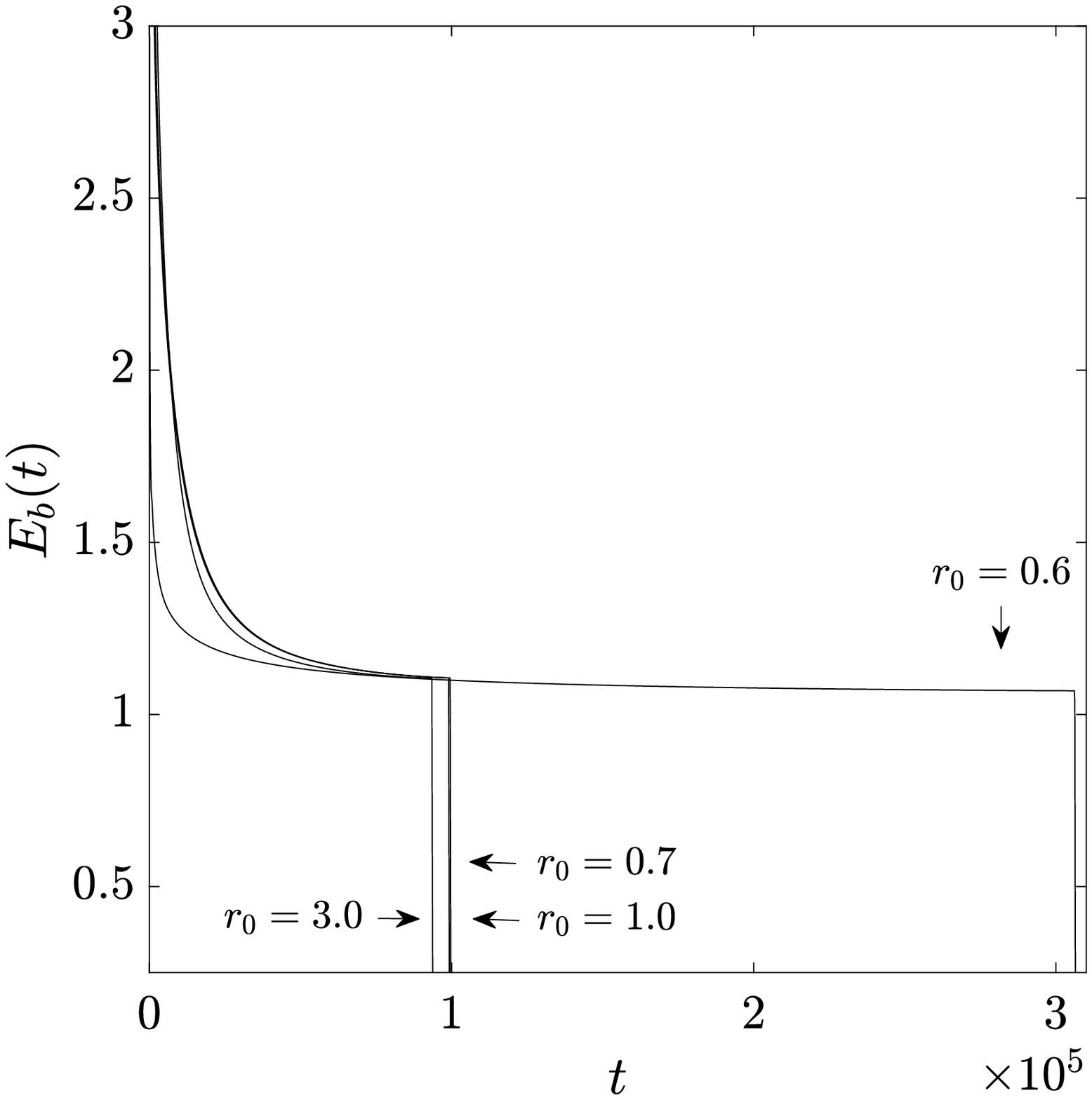}
	\caption{Time evolution of the bubble energy, $E_b(t)$ (cf. Eq. (\ref{eq24})), for the compacton potential (\ref{eq1}) for $n=3$ and distinct values of the bubble's radius $r_0$. We have used the resolution $N=700$ and $r_{abs}=40$.}
\end{figure}

We start with the compacton potential (\ref{eq1}). Figs. 5 and 6 show the plots of the bubble energy for $n=2$ and $n=3$, respectively, with the respective values of the radius $r_0$ indicated. We noticed the three typical stages consistent with the emergence of oscillons in the case $n=1$. In the first stage, the bubble release most of its energy. It is followed by a stage of approximately constant energy represented by a plateau with the scalar field undergoing oscillations in a finite region. During the last stage of evolution, the scalar field has its amplitude decreasing, and eventually, the oscillon radiates away its remaining energy. Apart from these features that effectively attest to oscillon configurations' existence, we highlight two distinct aspects. The first aspect is a large amount of bubbles' energy released in the first stage, increasing with $n$. More precisely, for $n=2$, we have about $29.78\%$ of the initial bubble energy in the oscillon configuration, while for $n=3$, the oscillon configuration has only about $17.76\%$. For reference, we have considered evaluating the bubble initial energy corresponding to the value of $r_0$ that produces the longest living oscillon phase and summarized the results in Table 1. For the sake of illustration, we present in Fig. 7 the phase diagram for the oscillon with $n=2$ and $r_0=94$. There are indications of the time duration of each phase of the oscillon evolution.

The second and more significant feature is the extended amount of duration of the oscillon, which are $t_{max} \sim 1.8 \times 10^5$ and $t_{max} \sim 3.0 \times 10^5$, for $n=2,3$, respectively. Notice that these values are much larger than the longest-lived oscillon of the $\phi^4$ model ($n=1$), $t_{max} \sim 9.9 \times 10^3$. For comparative purposes, we can see from Table 1 that for $n=1$ nearly $52.33\%$ of the initial bubble energy is stored in the oscillon phase.

\begin{table}
	\centerline{Table 1}
	\medskip
	\begin{center}
		\centering
		\begin{tabular}[c]{ l | c }
			\hline
			\\
			\textit{n} &$\frac{E_b}{E_{\mathrm{init}}} \times 100\, (\%)$  \\
			\\
			\hline
			\hline
			%& &\\
			$1\,(\phi^4\; model)$ & $52.33$  \\
			%& & \\
			%\vspace{0.2cm}
			%\hline
			$2$ & $29.78 $ \\
			%&  \\
			%\hline
			%\\
			$3$ & $17.76 $  \\
			$4$ & $7.75 $  \\
			\hline
		\end{tabular}
		\caption{Approximate amount of energy stored in the oscillon phase for $n=1,2,3$ and 4 for the compacton potential (\ref{eq1}).}
	\end{center}
\end{table}

The numerical simulations indicate clearly that the increase of $n$ can produce an oscillon configuration settled with a smaller fraction of the initial bubble energy. We show in Fig. 8 the time evolution of the bubble energy for $n=4$,  where the oscillon phase has retained approximately $7.75\%$ of the bubble's initial energy. In this case, $t_{max} \sim 2.5 \times 10^5$ for $r_0=0.7$.  

\begin{figure}[htb]
	\includegraphics[width=7cm,height=6cm]{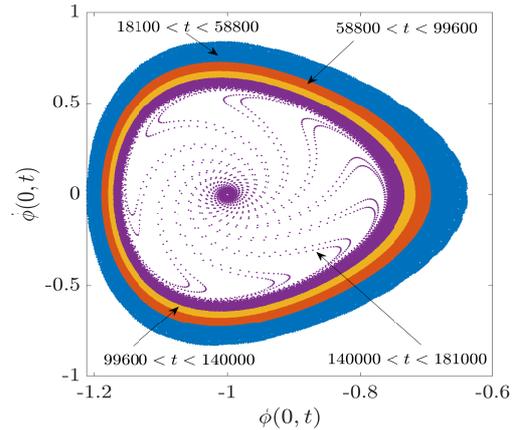}%{new_fig_7_r0_92_n_2}
	\caption{Phase space portrait of the oscillon evolution for $n=2$ (cf. Eq. (\ref{eq1})) and $r_0=0.94$. For clarity, we have depicted phases with approximately the same time interval in distinct colors. We also assumed the sampling at every $\Delta t=10$.}
\end{figure}

\begin{figure}[htb]
	\includegraphics[width=6.5cm,height=5.5cm]{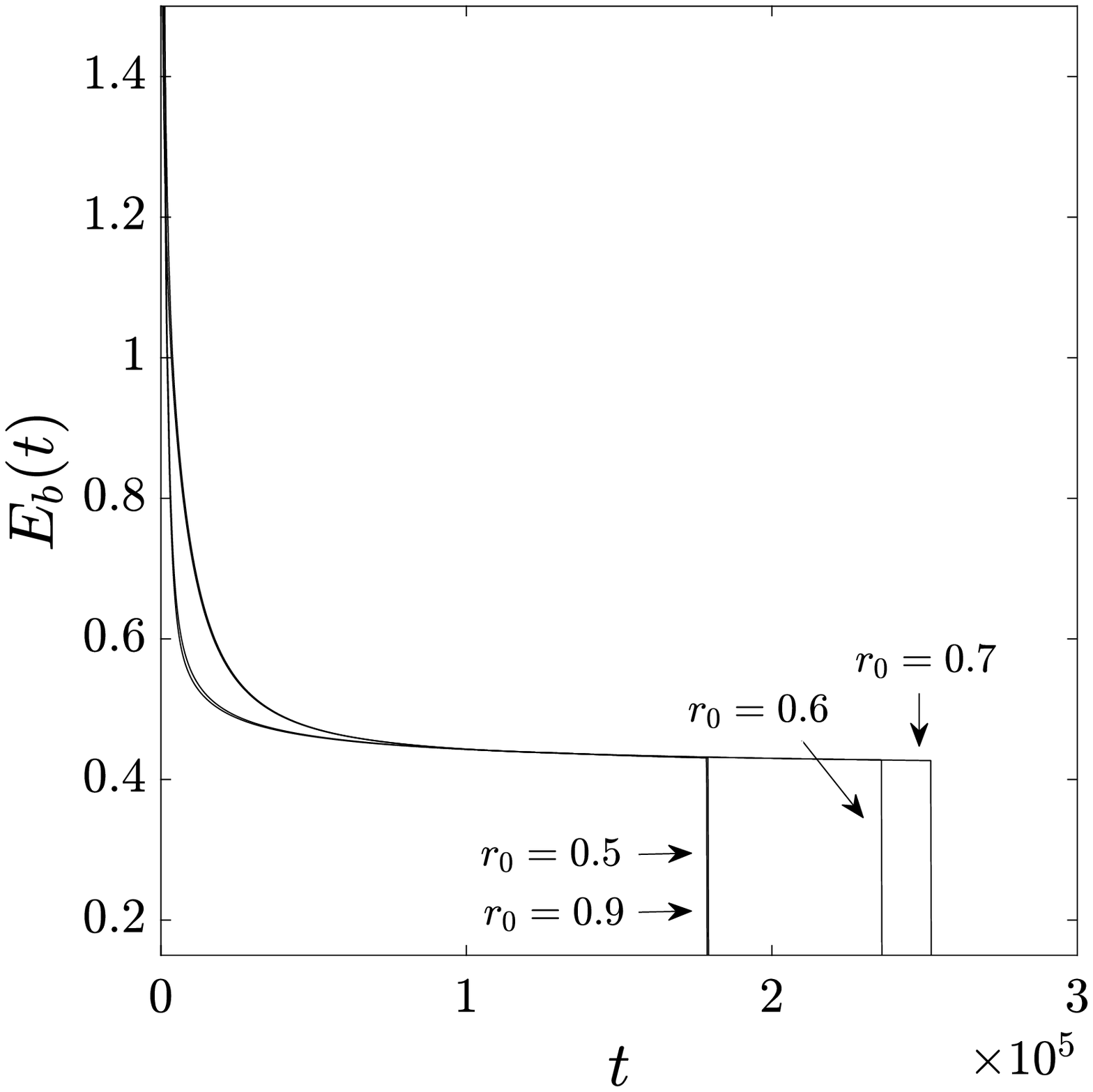}%{new_fig_n4}
	\caption{Time evolution of the bubble energy, $E_b(t)$ (cf. Eq. (\ref{eq24})), for the compacton potential (\ref{eq1}) for $n=4$ and distinct values of the bubble's radius $r_0$. We have used the resolution $N=750$ and $r_{abs}=15$.} 
\end{figure}

We further remark that in all cases, $n=2,3,4$, there exists a minimum value of the bubble radius for the scalar field to undergo in the oscillon configuration. However, we have not searched for the expected maximum radius as expected, but it seems to encompass relatively large values compared with the $\phi^4$ oscillons. 

\begin{figure}[htb]
	\includegraphics[width=6.5cm,height=5.5cm]{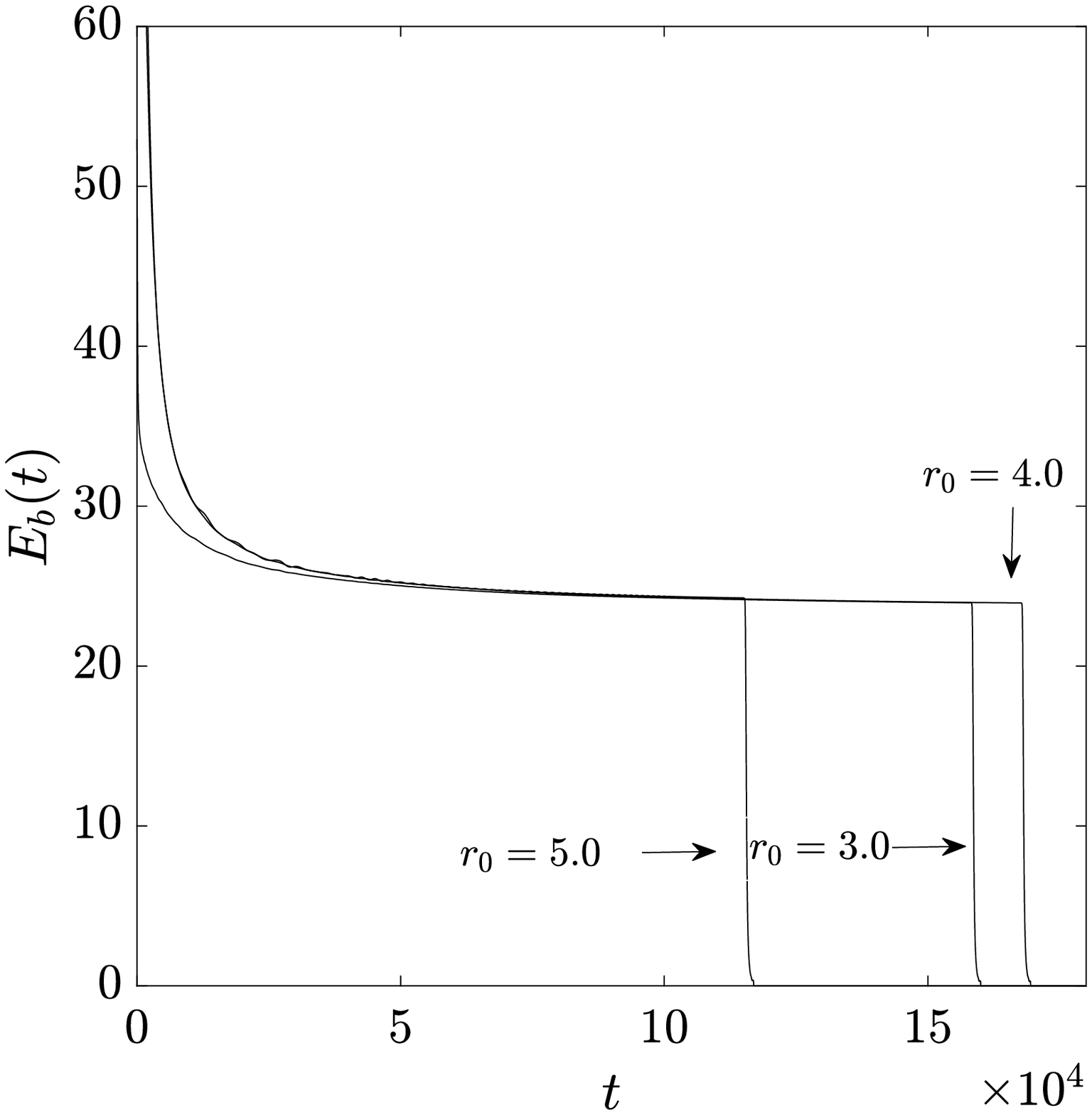}%{new_fig_phi6_energy_black}
	\caption{Time evolution of the bubble energy, $E_b(t)$ (cf. Eq. (\ref{eq24})), for the potential (\ref{eq2}) for $\epsilon=1$ and distinct values of the bubble's radius $r_0$. We have used the resolution $N=600$ and $r_{abs}=30$.} 
\end{figure}

Another long-lived oscillon phase is obtained with the modified $\phi^6$ model potential (cf. Eq. (\ref{eq2})),  where we have fixed $\epsilon=1$. We depict in Fig. 9 the time evolution of the bubbles' energies for several values of $r_0$. The plots reveal a long oscillon phase with about $54\%$ of the initial bubble energy lasting to $t_{max} \sim 1.7 \times 10^5$. For the sake of completeness, we present in Fig. 10 the scalar field at the origin or the bubble's core and the evolution of the phase diagram for the oscillon with $r_0=4$ and starting at $t \sim 1.7 \times 10^5$. The colors indicate distinct oscillon's radius during its lifetime.
 
\begin{figure*}[htb]
%\hspace{-0.7cm}	
\includegraphics[width=6cm,height=5.cm]{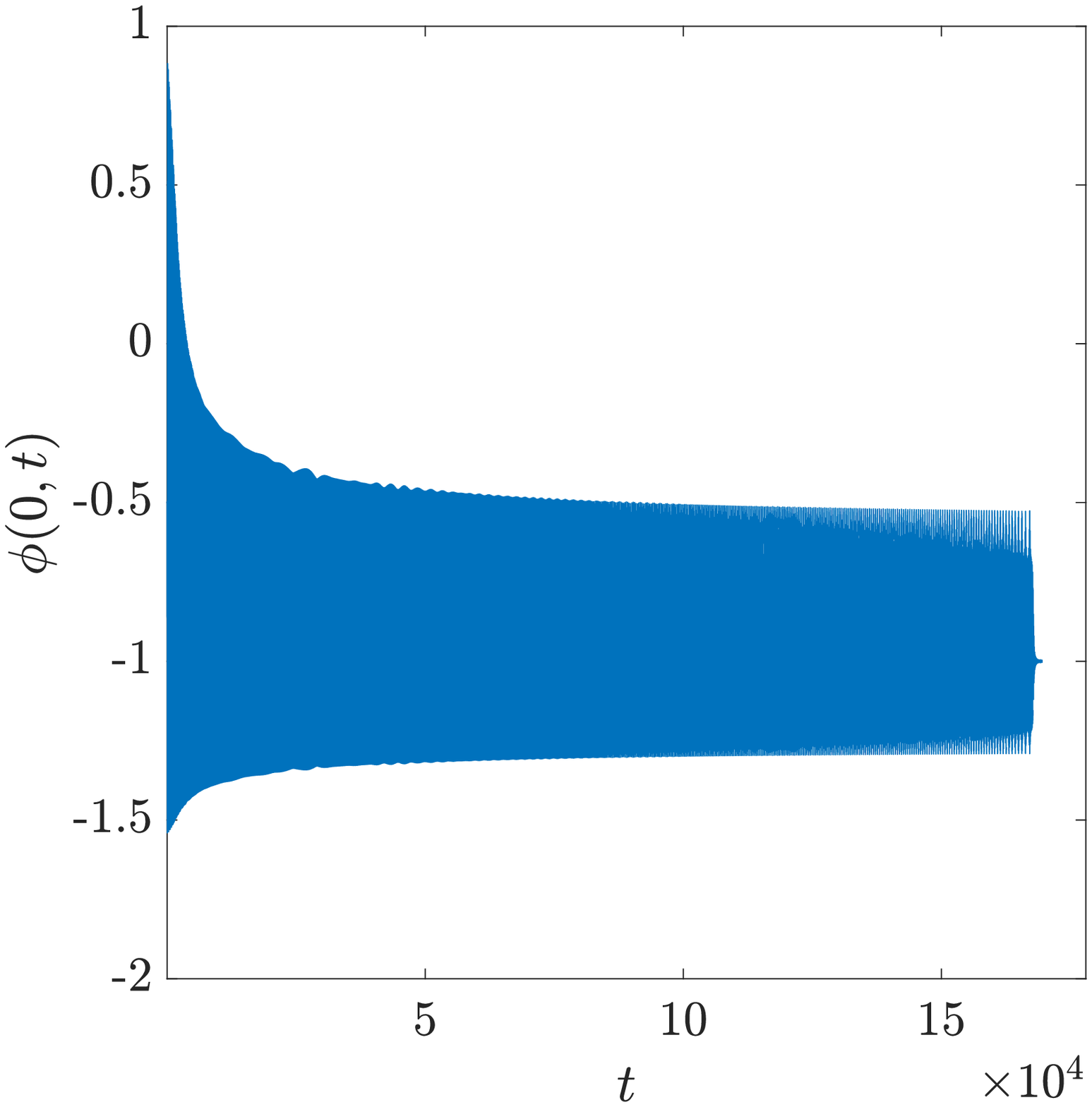} \hspace{1cm}
	\includegraphics[width=7cm,height=6cm]{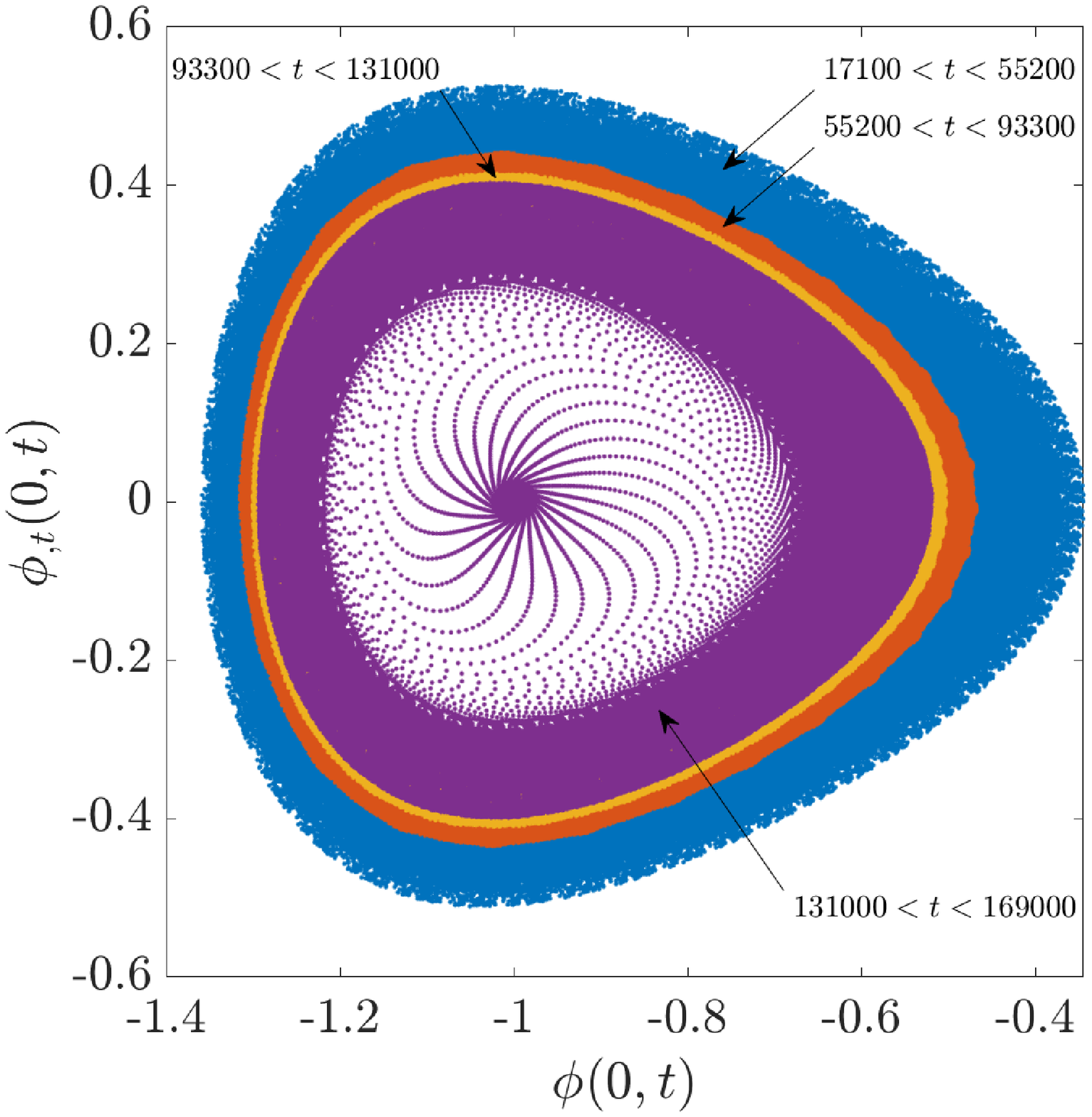} \\
	 (a) \hspace{7cm} (b)
	\caption{(a) Time evolution of the scalar field at the origin or the bubble's core for the parametric potential $\phi^6$ with $\epsilon=1$, $r_0=4$ and $L_0=20$. (b) Phase space portrait depicting phases with approximately the same time interval in distinct colors. We assumed the sampling at every $\Delta t=1$.}
	%\caption{(a) Time evolution of the scalar field at the origin or the bubble's core for the parametric potential $\phi^6$ with $\epsilon=1$, $r_0=4$ and $L_0=20$. (b) Phase space portrait depicting distinct phases of the oscillon evolution. We assumed the sampling at every $\Delta t=1$.} 
\end{figure*}

\section{Summary and Conclusions}

%In this paper, we present convincing numerical evidence of very long-living $3+1$ spherically symmetric oscillons. We have considered two different potentials instead of the traditional double-well potential of the $\phi^4$ model. The first is the compacton potential (cf. Eq. (\ref{eq1})) which describes $1+1$ dimensional topological defects with compact support as a generalization of the kinks of the $\phi^4$ model. The second is a modified or parametric $\phi^6$ potential proposed originally in Ref. \cite{christ_lee}. Both potentials belong to higher-order field theories.

This paper presents convincing numerical evidence of very long-living $3+1$ spherically symmetric oscillons. We have focused on two different potentials instead of the traditional double-well potential of the $\phi^4$ model. The first is the compacton potential (cf. Eq. (\ref{eq1})), which describes $1+1$ dimensional topological defects with compact support as a generalization of the kinks of the $\phi^4$ model. The second is a modified or parametric $\phi^6$ potential proposed originally in Ref. \cite{christ_lee}. Both potentials belong to higher-order field theories. However, for highly long-lived oscillons belonging to different potentials, see Refs. \cite{zhang, salmi,olle}.

We have implemented a Galerkin-Collocation spectral code to evolve the Klein-Gordon equation for extended time intervals. The novelty here is to use a set of basis functions in which each component satisfies the even parity condition about the origin. The algorithm proved to be exponentially convergent, and, as a qualitative validation test, we reproduced most of the results of Ref. \cite{gleiser_95}. 

The main result, however, is to show the existence of oscillon phases with highly long lifetimes about $t \sim \mathcal{O}(10^5)$ without fine-tuning the parameter $r_0$ (cf. Eq. (\ref{eq20})). We observed that by increasing $n$ present in the compacton potential, more energy is radiated away before the oscillon phase begins, as indicated in Table 1. On the other hand, oscillons generated with the $\phi^6$ potential retain about $54\%$ of the initial bubble energy. It is approximately the same as the oscillons of the $\phi^4$ model. 

We indicate some directions related to the present work. It is clear the relation between the radial parameter $r_0$ and the life span of the oscillon phase. We believe in the existence of a resonant structure as described in Ref. \cite{honda_choptuik}, albeit not investigate it thoroughly. Therefore, we can find oscillons with a more significant duration. Another feature of interest is the virialization \cite{gleiser_95} associated with the longer lifetime of the oscillon. The $2d$ oscillons is another possibility worth investigating, where we have preliminarily examined the potential (\ref{eq1}) with $n=2$ with the oscillon phase lasting more than $t \sim 10^6$ for a suitable value of $r_0$. Finally, the study of oscillons in curved spacetimes with the new potentials (\ref{eq1}) and (\ref{eq2}) would be an interesting generalization of the study of Ref. \cite{ikeda}.

\acknowledgments

H. P. O. acknowledges the financial support of the Brazilian Agency CNPq and the hospitality of the Department of Physics and Astronomy of the Bowdoin College. This study was financed in part by the Coordena\c c\~ao de Aperfei\c coamento de Pessoal de N\'ivel Superior - Brasil (CAPES) - Finance Code 001.  

\appendix*

\section{}

We present the procedure to calculate the scalar field energy given by Eq. (20), the integration with the dissipative term that appears in the energy balance equation (21), and the bubble's energy (cf. Eq. (25)).

We employ the Curtis-Clenshaw formula to approximate all integrals. The first step is to map the integration over the entire spatial domain $0 \leq r  < \infty$ into a computational domain labeled by $-1 \leq \xi \leq 1$ using the algebraic map

\[r = L_0 \frac{(1+\xi)}{1-\xi},\]

\noindent where $L_0$ is the map parameter. Then, it follows that

\begin{eqnarray}
E &=& 4\pi\,\int_{-1}^1\,\left(\frac{1}{2}\Pi^2+\frac{1}{2}\phi_{,x}^2\left(\frac{\partial x}{\partial r}\right)^2 + V(\phi)\right)r(\xi)^2\,\times \nonumber \\
\nonumber \\
&&\frac{dr}{d\xi}d\xi \approx \sum_{k=0}^{N_q}\,\left(...\right)_k w_k. %\label{eq22}
\end{eqnarray}

\noindent Here $N_q=2N$ is the quadrature truncation order, $\left(...\right)_k$ means the integrand evaluated at the quadrature Chebyshev-Gauss-Lobatto collocation points $\xi_k$ given by 

\[\xi_k = \cos\left(\frac{\pi k}{N_q}\right),\]

\noindent for all $k=0,1,..,N_q$. The weights are

\begin{eqnarray}
w_0&=&w_{N_q}=\frac{1}{N_q^2-1} \\
\nonumber \\
w_k &=& \frac{4}{N_q}\sum_{j=0}^{N_q/2}\,\cos\left(\frac{2\pi j  k}{N_q}\right)\frac{1}{c_j(1-4j^2)},
\end{eqnarray}

\noindent where $c_0=c_{N_q/2}=2$, and $c_j=1$ for $j=1,2,..,N_q/2-1$.

With the values of energy evaluated at each instant from Eq. (A1), we can calculate its time derivative using finite differences:

\begin{eqnarray}
\left(\frac{dE}{dt}\right)_{t_j} = \frac{1}{2 \delta t_j}\left[E(t_j+\delta t_j)-E(t_j-\delta t_j)\right]. 
\end{eqnarray}

\noindent where $\delta t_j=t_{j+1}-t_j$ is the stepize.

For the evaluation of the bubble's energy, we have first mapped the integration interval $0 \leq r < r_s$ into $-1 \leq \xi \leq 1$ with a linear map, $r=r_s(1+\xi)/2$, and applied the Curtis-Clenshaw formula straigth forwardly as presented.

\end{document}